\input{aipcheck}

\documentclass[
    ,final            
  ]
  {aipproc}

\layoutstyle{8x11single}
\usepackage{graphicx}
\usepackage{amsbsy}
\usepackage{bm}

\begin{document}

\title{Radiation from relativistic jets in turbulent magnetic fields}


\author{K.-I. Nishikawa}{
  address={National Space Science and Technology Center,
  Huntsville, AL 35805, USA}
}

\author{M. Medvedev}{
  address={Department of Physics and Astronomy, University of Kansas, KS
66045, USA}
}

\author{B. Zhang}{
  address={Department of Physics, University of Nevada, Las
Vegas, NV 89154, USA} }

\author{P. Hardee}{
  address={Department of Physics and Astronomy,
  The University of Alabama,
  Tuscaloosa, AL 35487, USA} }

\author{J. Niemiec}{
address={Institute of Nuclear Physics PAN, ul. Radzikowskiego 152, 31-342 Krak\'{o}w, Poland}}

\author{\AA. Nordlund}{
address={Niels Bohr Institute, University of Copenhagen, 
Juliane Maries Vej 30, 2100 Copenhagen \O, Denmark}
}

\author{J. Frederiksen}{
 address={Niels Bohr Institute, University of Copenhagen, 
Juliane Maries Vej 30, 2100 Copenhagen \O, Denmark}
}

\author{Y. Mizuno}{
  address={National Space Science and Technology Center,
  Huntsville, AL 35805, USA}}

\author{H. Sol}{
  address={LUTH, Observatore de Paris-Meudon, 5 place Jules Jansen, 92195 Meudon Cedex, France}}

\author{G. J. Fishman}{
  address={NASA/MSFC,
  Huntsville, AL 35805, USA} }

\begin{abstract}
%
Using our new 3-D relativistic electromagnetic particle (REMP) code
parallelized with MPI, we have investigated long-term particle
acceleration associated with an relativistic electron-positron jet
propagating in an unmagnetized ambient electron-positron plasma. The
simulations have been performed using a much longer simulation system
than our previous simulations in order to investigate the full
nonlinear stage of the Weibel instability and its particle
acceleration mechanism. Cold jet electrons are thermalized and ambient
electrons are accelerated in the resulting shocks. The acceleration of
ambient electrons leads to a maximum ambient electron density three
times larger than the original value. Behind the bow shock in the jet
shock strong electromagnetic fields are generated.  These fields may
lead to the afterglow emission.  We have calculated the time evolution
of the spectrum from two electrons propagating in a uniform parallel
magnetic field to verify the technique.
%
\end{abstract}

\maketitle

\vspace*{-0.8cm}
\section{RPIC simulations}
\vspace*{-0.2cm}
Particle-in-cell (PIC) simulations can shed light on the physical
mechanism of particle acceleration that occurs in the complicated
dynamics within relativistic shocks.  Recent PIC simulations of
relativistic electron-ion and electron-positron jets injected into an
ambient plasma show that acceleration occurs within the downstream jet
\cite{nishi03,nishi05,hede05,nishi06,ram07}.
%
%
In general, these  simulations have confirmed that
relativistic jets excite the Weibel instability. 
The Weibel instability generates current filaments and associated
magnetic fields \cite{medv99},
and accelerates electrons \cite{nishi03,nishi05,hede05,nishi06,ram07}.


\vspace{0.3cm}
\noindent
 {\bf  Pair Jets
Injected into Unmagnetized Pair  Plasmas using a Large System}
\vspace{0.1cm}

We have performed simulations using a system with ($L_{\rm x}, L_{\rm
y}, L_{\rm z}) = (4005\Delta, 105\Delta, 105\Delta)$ and a total of
$\sim 1$ billion particles (12 particles$/$cell$/$species for the
ambient plasma) in the active grid zones \cite{nishi08a}.  In the
simulations the electron skin depth, $\lambda_{\rm ce} = c/\omega_{\rm
pe} = 10.0\Delta$, where $\omega_{\rm pe} = (4\pi e^{2}n_{\rm
e}/m_{\rm e})^{1/2}$ is the electron plasma frequency and the electron
Debye length $\lambda_{\rm e}$ is half of the grid size. Here the
computational domain is six times longer than in our previous
simulations \cite{nishi06,ram07}.  The electron number density of the
jet is $0.676n_{\rm e}$, where $n_{\rm e}$ is the ambient electron
density and $\gamma = 15$.  The electron/positron thermal velocity of the
jet is $v^{\rm e}_{\rm j,th} = 0.014c$, where $c = 1$ is the speed of
light. Jets are injected in a plane across the computational grid at
$x = 25\Delta$ in the positive $x$ direction in order to eliminate
effects associated with the boundary conditions at $x = x_{\rm
\min}$. Radiating boundary conditions were used on the planes at {\it
$x = x_{\min}~{\&}~x_{\max}$}. Periodic boundary conditions were used
on all transverse boundaries.  The ambient and jet electron-positron
plasma has mass ratio $m_{\rm e^-}/m_{\rm e^+} = 1$.
%
The electron/positron thermal velocity in the ambient plasma is 
$v^{\rm e}_{\rm a,th} = 0.05c$. 

Figure 1 shows the averaged (in the $y-z$ plane) electron density and
electromagnetic field energy along the jet at $t = 2000\omega_{\rm
pe}^{-1}$ and $3750\omega_{\rm pe}^{-1}$.  The resulting profiles of
jet (red), ambient (blue), and total (black) electron density are
shown in Fig. 1a.  The ambient electrons are accelerated by the jet
electrons and pile up towards the front part of jet.  At the earlier time
the ambient plasma density increases linearly behind the jet front as shown by
the dashed blue line in Fig. 1a. At the later time the ambient plasma
shows a rapid increase to a plateau behind the jet front, with
additional increase to a higher plateau farther behind the jet front.
The jet density remains approximately constant except near the jet
front.

%
%
\vspace{-0.cm}
\begin{figure}[ht]
\begin{minipage}[t]{80mm}
\includegraphics[width=150pt, height=120pt]{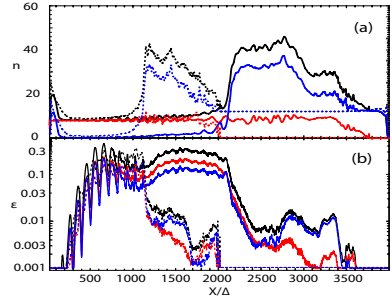}
\end{minipage}
\begin{minipage}[t]{50mm}
\vspace*{-12.0cm}
\caption{The averaged values of electron density (a) and field energy
(b) along the $x$ at $t = 3750\omega_{\rm pe}^{-1}$ (solid lines) and
$2000\omega_{\rm pe}^{-1}$ (dashed lines). Fig. 1a 
shows jet electrons (red), ambient electrons (blue),
and the total electron density (black). Fig. 1b shows electric field energy (blue),
magnetic field energy (red), and the total field energy (black) divided by the total kinetic energy.} 
\end{minipage}
\end{figure}
\vspace*{-0.4cm}

The Weibel instability remains excited by continuously injected
jet particles and the electromagnetic fields are kept at a high level,
about four times that seen in a previous much shorter grid
simulation system ($L_{\rm x} = 640\Delta$). At the earlier simulation
time ($t = 2000\omega_{\rm pe}^{-1}$) a large electromagnetic
oscillating structure is generated and accelerates the ambient
plasma. As shown in Fig. 1b, at the later simulation time the
oscillating structure extends up to $x/\Delta = 1100$, then becomes
more uniform and the magnetic field energy becomes larger than the
electric field energy. These strong electromagnetic fields become very
small beyond $x/\Delta = 2000$ in the shocked ambient region
\cite{nishi06,ram07}.

The acceleration of ambient electrons becomes visible when jet 
electrons pass about $x/\Delta = 500$. The maximum density of accelerated 
ambient electrons is attained at $t = 1750\omega_{\rm pe}^{-1}$.
The maximum density gradually reaches a plateau as seen in Fig. 1a. 
The maximum electromagnetic field energy is located at $x/\Delta = 700$ as shown in Fig. 1b. 
The location of this maximum remains in this region at large simulation times. 

\vspace{-0.5cm} \subsection{New Numerical Method of Calculating Synchrotron
and Jitter Emission from Electron Trajectories in Self-consistently
Generated Magnetic Fields}

\vspace{-0.1cm} 

Let a particle be at position ${\bf{r}_{0}}(t)$ at
time $t$  \cite{nishi08,hedeT05}. 
At the same time, 
we observe the electric field from the particle from position $\bf{r}$. However,
because of the finite propagation velocity of light, we observe the
particle at an earlier position $\bf{r}_{0}(\rm{t}^{'})$ where it was at
the retarded time $t^{'} = t - \delta t^{'} = t - \bf{R}(\rm{t}^{'})/c$. Here
$\bf{R}(\rm{t}^{'}) = |\bf{r} - \bf{r}_{0}(\rm{t}^{'})|$ is the distance from the
charge (at the retarded time $t^{'}$) to the observer.

After some calculation and simplifying assumptions the total energy
$W$ radiated per unit solid angle per unit frequency from a charged
particle moving with instantaneous velocity $\boldsymbol{\beta}$ under
acceleration $\boldsymbol{\dot{\beta}}$ can be expressed as

\vspace*{-0.5cm}
\begin{eqnarray}
\frac{d^{2}W}{d\Omega d\omega} & = & \frac{\mu_{0} c
q^{2}}{16\pi^{3}} \left|\int^{\infty}_{\infty}\frac{\bf{n}\times
[(\bf{n}-\boldsymbol{\beta})\times \boldsymbol{\dot{\beta}}]}{(1-\boldsymbol{\beta}\cdot
\bf{n})^{2}} e^{i\omega(t^{'} -\bf{n} \cdot \bf{r}_{0}({\rm t}^{'})/{\rm c})}
dt^{'}\right|^{2}
\end{eqnarray}
\vspace*{-0.3cm} 

Here, $\bf{n} \equiv \bf{R}(\rm{t}^{'})/ |\bf{R}(\rm{t}^{'})|$ is a
unit vector that points from the particle's retarded position towards
the observer. 
The choice of unit vector $\bf{n}$ along the direction of propagation of
the jet (hereafter taken to be the $x$-axis) corresponds to head-on emission. 
For any other choice of $\bf{n}$ (e.g., $\theta = 1/\gamma$), off-axis emission 
is seen by the observer. The observer's viewing angle is set by the choice of 
$\bf{n}$ ($n_{\rm x}^{2}+n_{\rm y}^{2}+n_{\rm z}^{2} = 1$). 

 
In order to calculate radiation from relativistic jets  propagating along the $x$ 
direction \cite{nishi08} we consider a test case which includes a parallel 
magnetic field ($B_{\rm x}$), and jet velocity of $v_{\rm j1,2} = 0.99c$. Two electrons 
are injected with different perpendicular velocities ($v_{\perp 1} = 0.1c, v_{\perp 2} 
= 0.12c$). A maximum Lorenz factor of 
$\gamma_{\max} =\{(1 - (v_{\rm j2}^{2} +v_{\perp 2}^{2})/c^{2}\}^{-1/2}  
= 13.48$ accompanies the larger perpendicular velocity.

\begin{figure}[h]
\includegraphics[width=110pt, height=110pt]{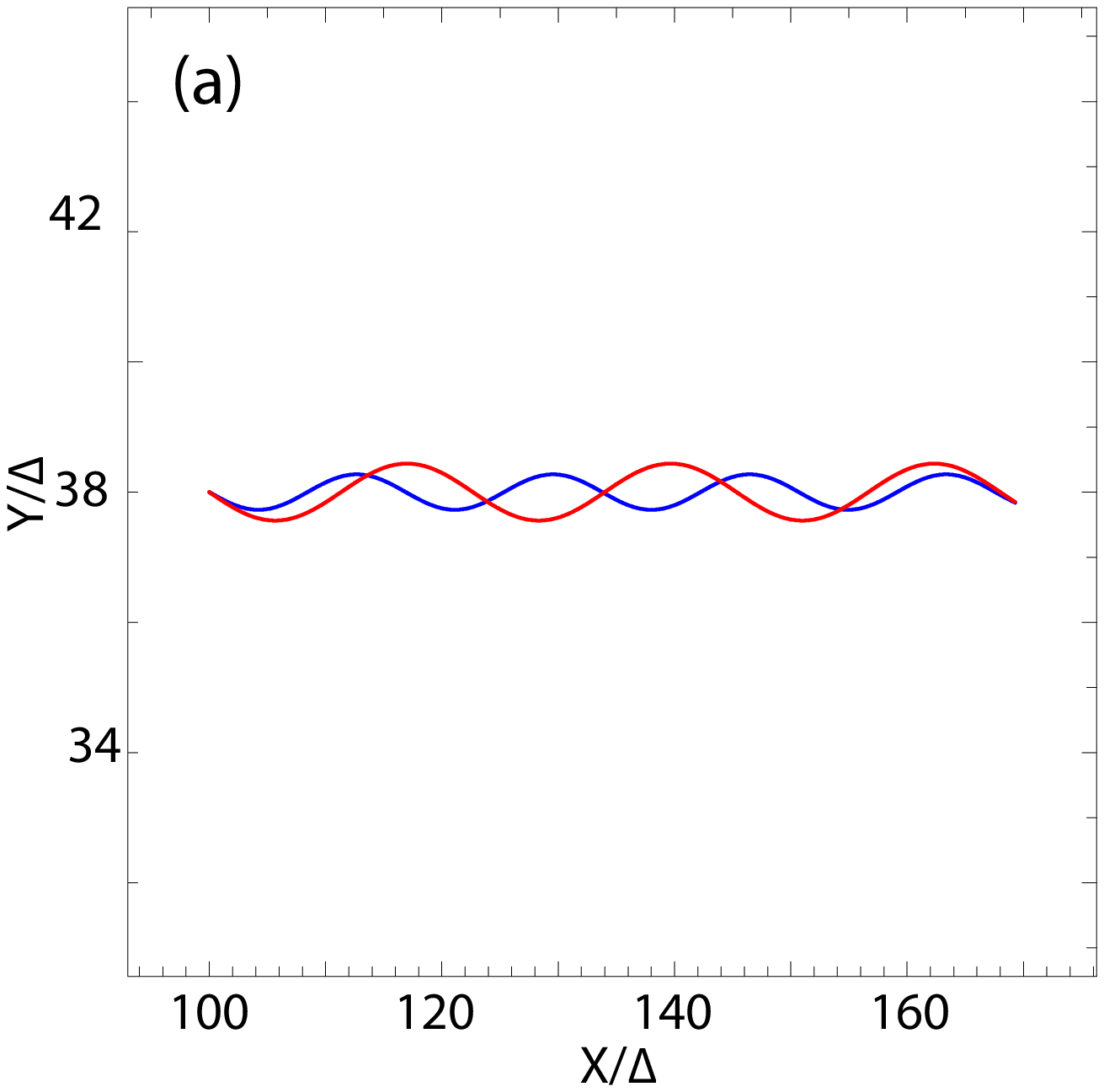}
\includegraphics[width=200pt, height=100pt]{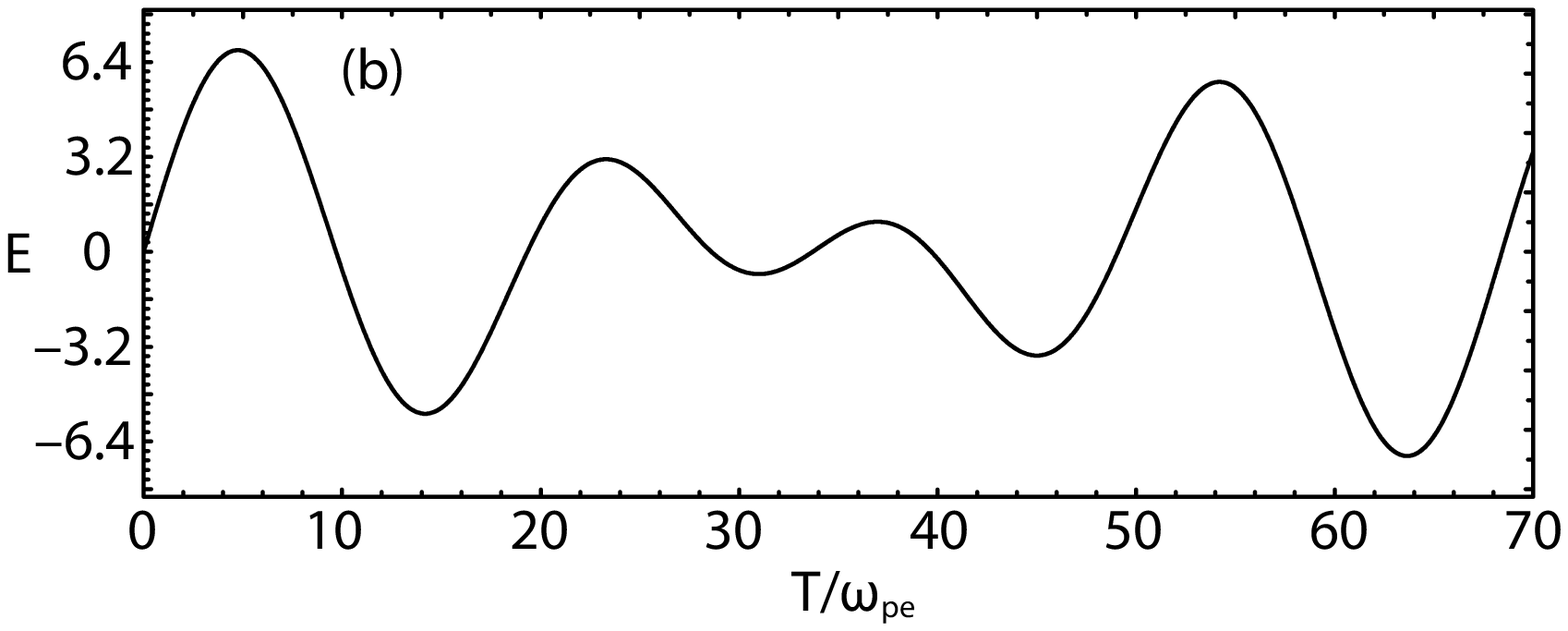}
\hspace*{0.0cm}
\includegraphics[width=120pt, height=120pt]{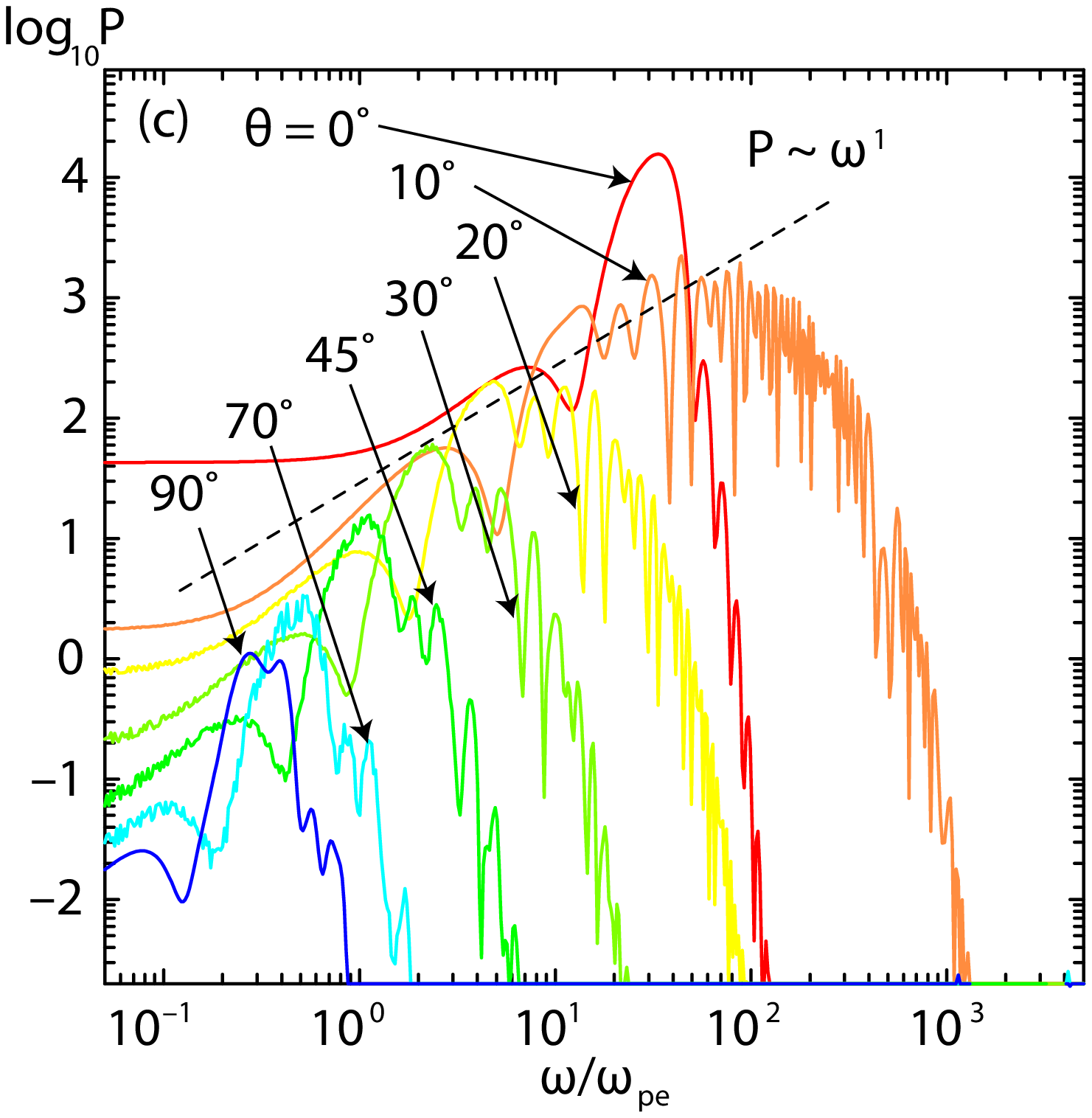}
\vspace*{-0.5cm}
\caption{\baselineskip 12pt The paths of two electrons moving helically along the $x-$direction
in a homogenous magnetic ($B_{\rm x}$)
field shown in the $x-y$-plane (a). The two electrons radiate a time dependent electric field. An observer
situated at great distance along the n-vector sees the retarded electric field from
the moving electrons (b). 
The observed power spectrum at different viewing angles from the two electrons (c). 
Frequency is in units of $\omega_{\rm pe}^{-1}$. }
\end{figure}

Figure 2 shows electron trajectories in the $x - y$ plane (a: left
panel), the radiation (retarded) electric field (red: $v_{\perp 1} =
0.12c$, blue: $v_{\perp 1} = 0.1c$) (b: middle panel), and spectra
(right panel) for the case $B_{\rm x} = 3.70$. The two electrons are
propagating left to right with gyration in the $y - z$ plane (not
shown). The gyroradius is about $0.44\Delta$ for the electron with the
larger perpendicular velocity. The power spectra were calculated at
the point $(x, y, z) = (64,000,000\Delta, 43.0\Delta,
43.0\Delta)$. The seven curves show the power spectrum at viewing
angles of 0$^{\circ}$ (red), 10$^{\circ}$ (orange), 20$^{\circ}$ (yellow),
30$^{\circ}$ (moss green), 45$^{\circ}$ (green), 70$^{\circ}$ (light
blue), and 90$^{\circ}$ (blue). The higher frequencies become stronger
at the $10^{\circ}$ viewing angle . The critical angle for off-axis
radiation $\theta_{\gamma} = \gamma_{\max}^{-1}$ for this case is
13.48$^{\circ}$. As shown in this panel, the spectrum at a larger
viewing angle ($> 20^{\circ}$) has smaller amplitude.

Since the jet plasma has a large velocity $x$-component in the
simulation frame, the radiation from the particles
(electrons and positrons) is heavily beamed along the $x$-axis
(jitter radiation) \cite{medv06}. 

In order to obtain the spectrum of synchrotron (jitter) emission, we 
consider an ensemble of electrons selected in the region where the
Weibel instability has grown fully and electrons are accelerated in
the generated magnetic fields. We will calculate emission from about
20,000 electrons during the sampling time $t_{\rm s} = t_{\rm 2} -
t_{\rm 1}$ with Nyquist frequency $\omega_{\rm N} = 1/2\Delta t$ where
$\Delta t$ is the simulation time step and the frequency resolution
$\Delta \omega = 1/t_{\rm s}$. 


Emission obtained with the method described above is self-consistent,
and automatically accounts for magnetic field structures on the
small scales responsible for jitter emission. By performing such
calculations for simulations with different parameters, we can 
investigate and compare the quite different regimes of jitter- and
synchrotron-type emission \cite{medv06}. 
The feasibility
of this approach has already been demonstrated \cite{hedeT05,hedeN05}, 
and its implementation is straightforward.
Thus, we should be able to address the low frequency GRB spectral
index violation of the synchrotron spectrum line of death \cite{medv06}. 


\vspace*{-0.3cm}
\begin{theacknowledgments}
\vspace*{-0.2cm}
This work is supported by AST-0506719, 
AST-0506666, NASA-NNG05GK73G, NNX07AJ88G, NNX08AG83G, NNX08AL39G, and NNX09AD16G.
 JN was supported by MNiSW research 
projects 1 P03D 003 29 and N N203 393034,
 and The Foundation for Polish Science through the HOMING program, which is
 supported through the EEA Financial Mechanism.Simulations were performed at the  
Columbia facility at the NASA
Advanced Supercomputing (NAS).  and IBM p690 (Copper) at the National
Center for Supercomputing Applications (NCSA) which is supported by
the NSF. 
Part of this work was done while K.-I. N. was visiting the
Niels Bohr Institute. Support from the Danish Natural Science Research Council is gratefully acknowledged.

\end{theacknowledgments}



\bibliographystyle{aipproc}   




\end{document}